\documentclass[conference]{IEEEtran}
\pdfoutput=1
\IEEEoverridecommandlockouts
\usepackage{cite}
\usepackage{amsmath,amssymb,amsfonts}
\usepackage{algorithmic}
\usepackage{graphicx}
\usepackage{hyperref}
\usepackage{subfiles}
\usepackage{gensymb}
\usepackage{subfigure}

\usepackage{textcomp}
\usepackage{xcolor}
\usepackage{amsmath}
\usepackage{gensymb}
\usepackage{graphicx}
\usepackage{caption}

\usepackage{etoolbox}

\begin{document}

\title{Modeling the Impact of 5G Leakage on Weather Prediction\\
\thanks{This work is supported in part by the NSF under Grant No. ACI-1541069.}
}

\author{\IEEEauthorblockN{ Mohammad Yousefvand}
\IEEEauthorblockA{\textit{WINLAB, Department of ECE } \\
\textit{Rutgers University}\\
North Brunswick, NJ, USA \\
my342@winlab.rutgers.edu}
\and
\IEEEauthorblockN{Chung-Tse Michael Wu}
\IEEEauthorblockA{\textit{Microwave Lab, Department of ECE} \\
\textit{Rutgers University}\\
Piscataway, NJ, USA \\
ctm.wu@rutgers.edu}
\and
\IEEEauthorblockN{Ruo-Qian Wang}
\IEEEauthorblockA{\textit{WHIR Lab, Department of CEE} \\
\textit{Rutgers University}\\
Piscataway, NJ, USA \\
rq.wang@rutgers.edu}
\and 
\IEEEauthorblockN{Joseph Brodie}
\IEEEauthorblockA{\textit{Rutgers Center for Ocean Observing Leadership} \\
\textit{Rutgers University}\\
New Brunswick, NJ, USA \\
jbrodie@marine.rutgers.edu}
\and
\IEEEauthorblockN{Narayan Mandayam}
\IEEEauthorblockA{\textit{WINLAB, Department of ECE} \\
\textit{Rutgers University}\\
North Brunswick, NJ, USA \\
narayan@winlab.rutgers.edu}
}

\maketitle

\begin{abstract}
The 5G band allocated in the 26 GHz spectrum referred to as 3GPP band n258, has generated a lot of anxiety and concern in the meteorological data forecasting community including the National Oceanic and Atmospheric Administration (NOAA). Unlike traditional spectrum coexistence problems, the issue here stems from the leakage of n258 band transmissions impacting the observations of passive sensors (e.g. AMSU-A) operating at 23.8 GHz on weather satellites used to detect the amount of water vapor in the atmosphere, which in turn affects weather forecasting and predictions. In this paper, we study the impact of 5G leakage on the accuracy of data assimilation based weather prediction algorithms by using a first order propagation model to characterize the effect of the leakage signal on the brightness temperature (atmospheric radiance) and the induced noise temperature at the receiving antenna of the passive sensor (radiometer) on the weather observation satellite. We then characterize the resulting inaccuracies when using the Weather Research and Forecasting Data Assimilation model (WRFDA) to predict temperature and rainfall. For example, the impact of 5G leakage of -20dBW to -15dBW on the well-known Super Tuesday Tornado Outbreak data set, affects the meteorological forecasting up to 0.9 mm in precipitation and 1.3 °C in 2m-temperature. We outline future directions for both improved modeling of 5G leakage effects as well as mitigation using cross-layer antenna techniques coupled with resource allocation.
\end{abstract}

\begin{IEEEkeywords}
5G, mmWave, weather prediction, leakage, n258 band, radiance.
\end{IEEEkeywords}

\section{Introduction}
Due to the shortage of available spectrum in sub-6 GHz frequency bands for cellular communications, mmWave frequency bands with large spectrum availability are considered in 5G to  enable cellular service providers to cope with the increasing demand for higher data rates and ultra low latency services \cite{Rappaport:2013:Millimeter}. The major 5G mmWave bands are 26 GHz (n258 band), 28 GHz (n257 band), 39 GHz (n260 band), and 47 GHz~\cite{Whitepaper:2019:5GMillimeterWaveFrequencies}. 
Of specific interest is the 3GPP band n258 band (see Figure 1), which is adjacent to 23.8 GHz where the passive sensors (e.g. Advanced Microwave Sounding Unit (AMSU)-A sensors \cite{AMSU}  embedded in weather prediction satellites operate to dynamically monitor and measure the atmospheric radiance which is used to predict the density of water vapor in the atmosphere, that is then further used in weather forecasting. The adjacency of the 23.8 GHz frequency, used by National Oceanic and Atmospheric Administration (NOAA) weather prediction satellites, to the n258 band used by 5G equipment results in inter channel interference which could negatively impact the precision of the underlying weather forecast models. In fact, the leakage of energy from the 5G bands into the 23.8 GHz band perturbs the radiance (equivalently brightness temperature) of atmospheric thermal emissions that is observed and measured by the passive sensors on the weather satellites, thereby lowering the validity and precision of weather forecast models. 
 
 Such coexistence and interdependence issues between 5G mmWave networks and weather prediction satellites raises concern and speculation over the potential negative impact of 5G services and radio transmissions on weather forecasting. In fact, the potential use of the n258 band has generated a lot of anxiety and concern in the meteorological data forecasting community including the National Oceanic and Atmospheric Administration (NOAA) \cite{Hollister:2019:5G, Witze:2019:Global,Segan:2019:5G,Kunkee2020}. Hence, such 5G leakage needs to be precisely characterized and addressed in order to maintain the accuracy of the satellite based weather forecasts \cite{online:2019:5GMesswithWeatherForecasts}. In fact, recent versions of the 3GPP’s 5G NR specification specifically have a carveout to protect satellite weather services, by reducing the emission levels of neighboring 5G signals between 24.25 and 27.5 GHz~\cite{ETSI}. But NOAA is arguing the current emission requirements aren't enough — it’ll lose that critical data required for precise forecasting unless such 5G emissions are clamped down even further. Understanding and characterizing the effect of such ``spectrum coexistence" calls for an interdisciplinary approach to first understand the impact of 5G transmissions on weather data measurements and prediction, and second, design mitigation strategies as needed to enable seamless coexistence between 5G services and weather prediction.
 
\begin{figure}[htb!]
	\includegraphics[width=\linewidth]{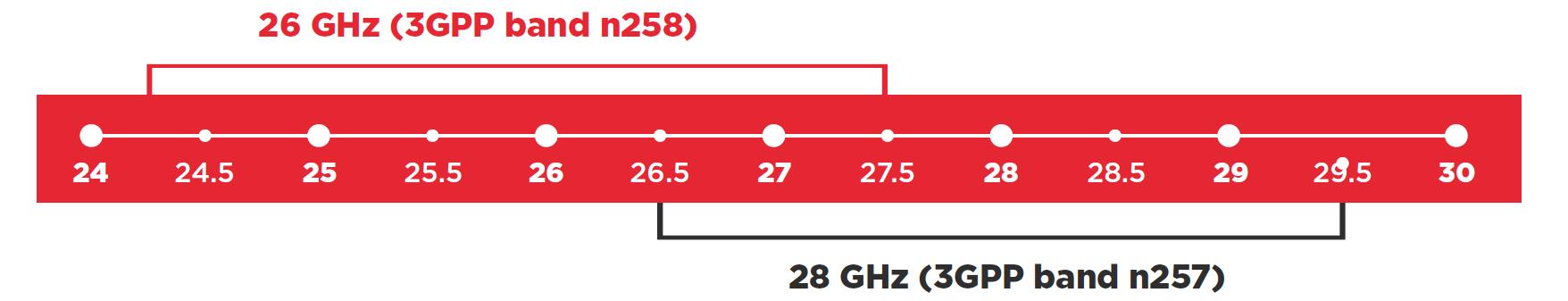}
	\centering
	\caption{5G Spectrum Allocations in 26 GHz and 28 GHz}
	\label{28GHz}
\end{figure}
In this paper, we study the impact of 5G leakage on the precision of data assimilation based numerical weather prediction (NWP) models~\cite{Liu:2012:ImpactofAssimilating,Schwartz:2019:RadianceDataAssimilation}. We begin by considering a first order propagation model to characterize the effect of the leakage signal on the brightness temperature (atmospheric radiance) and the induced noise temperature at the receiving antenna of the passive sensor (radiometer) on the weather observation satellite. We then characterize the effects of errors and inaccuracies in measured radiance/noise temperatures, as a result of 5G leakage, on the temperature and rainfall forecasts of NWP models. Finally, we provide simulations results based on the proposed model to show that the leakage from 5G transmissions could negatively impact the accuracy of ultimate temperature and rainfall predictions of NWP models, and  motivate the need for further research in this area.

The rest of this paper is organized as follows. Section~\ref{Sec2:5GLeakage} explains how 5G leakage could impact weather prediction. Section~\ref{sec3:DataAssimilation} introduces data assimilation based NWP models, in which the atmospheric radiance measurement is one of the major inputs for prediction. In section~\ref{sec4:NumericalResults}, we provide simulation results to show how radiance measurement error/perturbation as a result of 5G leakage could negatively impact the precision of weather forecast models. We then discuss several future directions for both improved modeling of 5G leakage effects as well as mitigating its negative impact on weather prediction in section~\ref{sec5:FutureDirections}, and conclude in section~\ref{sec6:Conclusion}. 

\section{Impact of 5G Leakage on the Radiance Measurements of Passive Sensors on Satellites}
\label{Sec2:5GLeakage}
\begin{figure}
    \centering
    \includegraphics[width=0.40\textwidth]{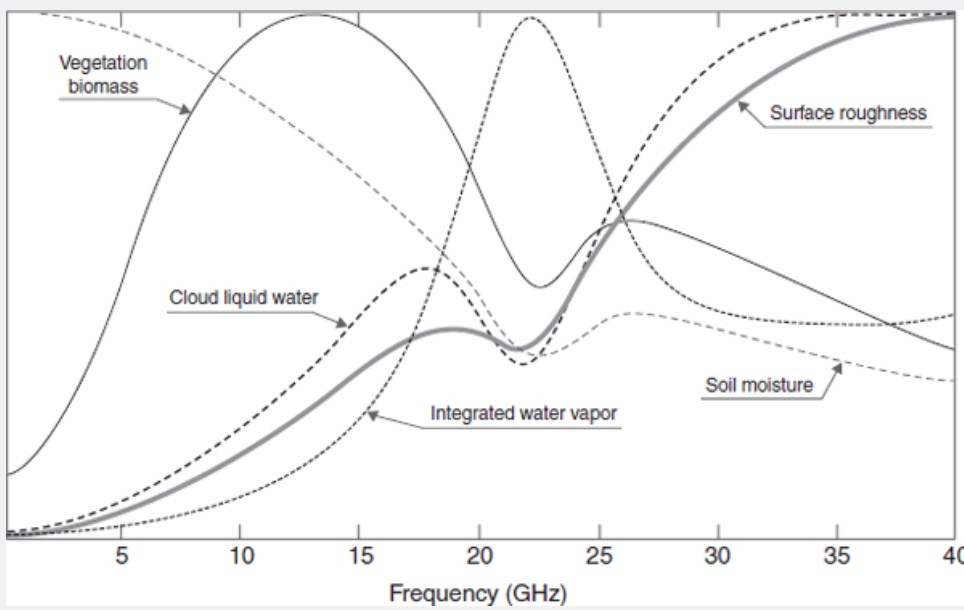}
    \caption{Absorption of EM Spectrum}
    \label{Absorption}
\end{figure}

The 5G transmissions will involve many frequencies, but the key one under discussion is the n258 band because of its proximity to the 23.8 GHz spectrum. The reason why the 23.8 GHz band plays a critical role in weather prediction is because of its importance in water vapor measurements in the atmosphere. Specifically, the absorption of electromagnetic signals in this band due to water vapor is much higher than other bands, and also the sensitivity of this band to other atmospheric factors is less than its sensitivity to vapor density (see Fig.~\ref{Absorption}), which makes this band an ideal band for water vapor measurements~\cite{online:2019:5GMesswithWeatherForecasts}.

Hence, when 5G equipment are transmitting signals in bands near the 23.8 GHz frequency (such as the n258 band), a passive sensor on a weather satellite measuring water vapor might be affected, resulting in erroneous data which in turn could degrade weather predictions that rely on accurate measurements of water vapor content in the atmosphere. The debate about what is considered as acceptable leakage from 5G transmissions ranges from a low of $-55$ $dBW$ to a high of $-20$ $dBW$~\cite{Witze:2019:Global}.

There are several sources of radiance in any given geographical area that contribute to the aggregate radiance measured by passive radiometers, operating in the 23.8 GHz band, on satellites located at a given altitude from the ground level. However, the radiance originating from the 5G transmissions in the adjacent n258 band are not currently considered and recognized as a major source of radiance in radiance models used for weather forecasting. Hence, the 5G leakage signal in the 23.8 GHz band will be likely perceived by the NWP models as an extra radiance originated from other sources like atmospheric emissions, and hence it will result in errors in measuring the water vapor density of the atmosphere in a given geographical area, which could in turn affect the underlying regional weather forecasts that rely on such measurements.

\begin{figure}[t!]
	\includegraphics[width=\linewidth]{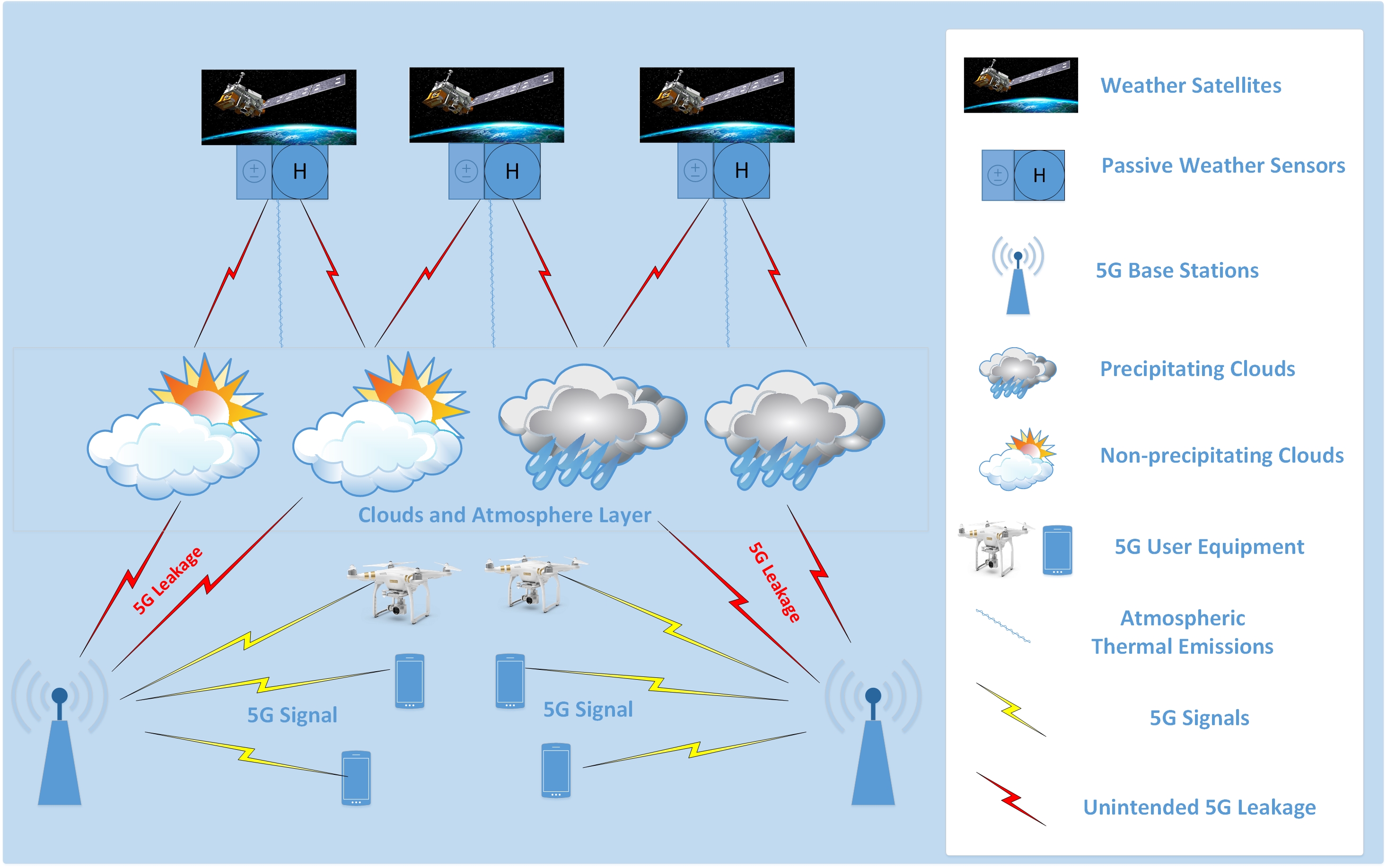}
	\centering
	\caption{5G Interference on Passive Sensors on Weather Observation Satellites}
	\label{Coexistence}
	\vspace{-0.1in}
\end{figure}

In fact, as illustrated in Fig.~\ref{Coexistence}, the leakage of energy from the 5G bands into the 23.8 GHz band perturbs the radiance (equivalently brightness temperature) of atmospheric thermal emissions that is observed and measured by the passive sensors on the weather satellites, thereby lowering the validity and precision of weather forecast models. 
As we can see in Fig. \ref{Coexistence}, the unintended 5G leakage signals will pass through the clouds and atmosphere layer, before reaching the radiometers at satellite sensors, during which parts of their energy will be absorbed by the particles in this layer, and the rest of their energy will be radiated out from this layer upward toward the observing weather satellites.
The intensity of the aggregate leakage signal resulted from 5G transmissions in the n258 band in a particular geographical area into the 23.8 GHz band, and accordingly its received power on the passive sensors of weather satellites, depends on a variety of factors such as the spatial density of 5G transmitters, elevation and directionality of transmissions, transmit power levels, specific sub-bands occupied, transmit modulation schemes chosen, the nonlinearity distortion of power amplifiers, and absorption and transmittance coefficients associated to the clouds and  atmosphere layer in which 5G signals are passing through.

\begin{figure}[htb!]
    \centering
    \includegraphics[width=\linewidth]{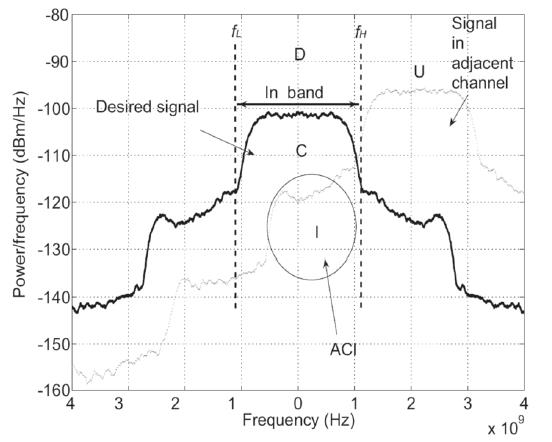}
    \caption{Adjacent Channel Interference (ACI).}
    \label{Leakage}
\end{figure}

\begin{figure} [htb!]
    \centering
    \includegraphics[width=\linewidth]{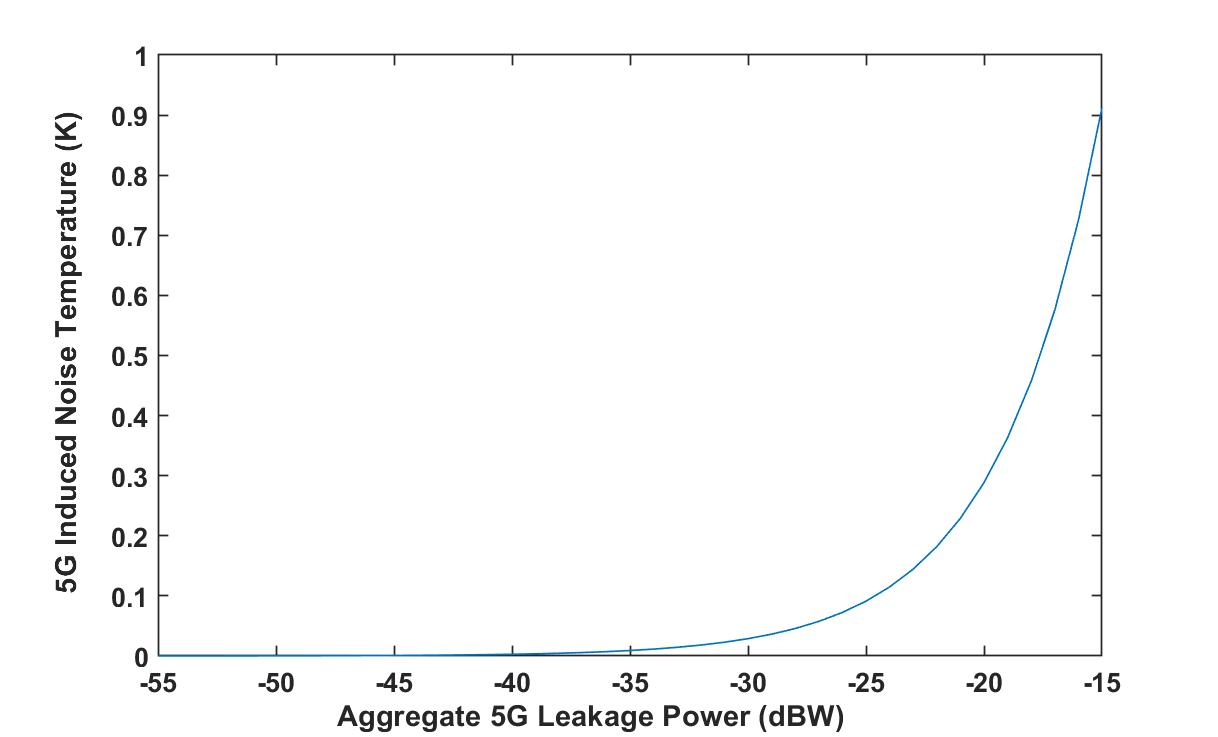}
    \vspace{-3mm}
    \caption{5G Induced Noise Temperature Associated with Different 5G Leakage Powers.}
    \label{NoiseFigure}
\end{figure}

To characterize the aggregate 5G leakage power, we use a preliminary first order analytical approach using an Adjacent Channel Interference (ACI) model~\cite{Ming:2008:ACI} along with a simplified pathloss model as follows. As shown in Fig. \ref{Leakage}, ACI seen by a particular channel $D$ is characterized by the generated out-of-band spectrum regrowth/leakage of an adjacent channel $U$, falling into the in-band region (between lower frequency $f_L$ and higher frequency $f_H$) of the desired channel $D$. 
For the AMSU-A passive sensors, the operating channel bandwidth is 270 MHz~\cite{AMSU}. Since the NOAA weather satellites instrumented with AMSU-A are on average between 700-900 km above the earth, we assume a nominal distance of 800 km and use a free space pathloss model to characterize the induced noise temperature due to 5G leakage at the receiving antenna of the passive sensor (radiometer) on the weather observation satellite. Note that the antenna noise temperature $T_a$ at the radiometers and the atmosphere brightness temperature (radiance) $T_b$ measured at the radiometers are linearly related to each other as
\begin{equation}
    T_a = \frac{T_b}{L} + \frac{L-1}{L} T_p = \eta_{rad} T_b + (1-\eta_{rad}) T_p,
\end{equation}
where $L$ is the loss factor, $T_p$ is the thermal noise temperature generated by the antenna, and $\eta_{rad}$ is the antenna radiation efficiency ~\cite{pozar2009microwave}. 
The additional noise temperature of radiometers induced by 5G signals will also affect the underlying brightness temperature measured by them. Using the expression $\frac{P_N}{B}= K_B T$, where $P_N$ is the induced noise power at the radiometer (in Watts), $B$ is the channel bandwidth (in Hertz) over which the noise power is measured, $K_B$ is the is the Boltzmann constant ($1.381 \times 10^{23}$ J/K, Joules per Kelvin), and $T$ is the noise temperature (in Kelvin), we show in  Fig. \ref{NoiseFigure} the 5G induced noise temperature at the radiometers as a function of different 5G leakage powers from terrestrial emissions of 5G equipment. In the figure, we have made an idealized assumption of negligible blockage and scattering of the leakage signal through the atmosphere and set the total pathloss to be 130 dB after appropriate antenna and system gains. In reality, the 5G leakage signals will likely pass through a layer of clouds and atmosphere before reaching the satellite sensors and parts of their energy will be absorbed by this layer \cite{Schwartz:2019:RadianceDataAssimilation}, hence, this absorption effect should be captured in improved propagation models as suggested in the section \ref{sec5:FutureDirections}. Using the first order analysis above, we next show the impact of 5G leakage on weather forecasts.

\section{Overview of data assimilation models used for weather prediction}
\label{sec3:DataAssimilation}
Many NWP models implement data assimilation (DA) algorithms, in which they use realtime observations obtained from sensors and monitoring instruments to update their state and boundary conditions to reduce the prediction error \cite{Liu:2012:ImpactofAssimilating}. In radiometry, radiance is defined as the flux density of radiant energy per unit solid angle of propagation and per unit projected area of radiating surface~\cite{RadianceDefinition}. Radiance observation data obtained from satellites are among the most important observation types that affect the performance of NWP models, and usually are assimilated in variational DA algorithms used in such models to improve the reliability and precision of forecasts~\cite{Liu:2012:ImpactofAssimilating}. Passive sensors (radiometers) operating in the 23.8 GHz band and housed on NOAA satellites located at altitudes ranging from around 700-900 km above the earth's surface measure the radiance of the earth's atmosphere. Such radiance is sensitive to the earth's atmospheric characteristics and atmospheric variables such as water vapor, concentration of trace gases, among many others. 

Radiance observations are prone to systematic errors (i.e., biases) that must be corrected before they are assimilated in NWP models. The radiance bias is often expressed by a linear combination of predictors, which leads to the definition of a modified forward operator $\hat{H}$ given by
\begin{equation}
\hat{H}(\underline{x_r}, \underline{\beta})=H(\underline{x})+ \beta_0 + \sum_{i=1}^{N_p}\beta_i * p_i,
\label{eq1}
\end{equation}
\noindent where $H(\underline{x})$ is the original forward operator, $\underline{x}$ is the model state vector, $\beta_0$ is the constant bias coefficient, $p_i$ is the value of the $i$-th predictor out of $N_p$ predictors, and $\beta_i$ is the bias correction coefficient associated to the $i$-th predictor \cite{Liu:2012:ImpactofAssimilating}. Predictors can be classified as those that are related to the model state, such as surface temperature and layer thickness, and those that are related to the measurement, such as measured radiations, or position of scan.

The bias-correction coefficients in vector $\underline{\beta}$ are updated iteratively within a variational minimization process. A common algorithm called 3DVar can be realized with the cost function $j(\underline{\beta})$ for the minimization
\begin{equation}
\begin{split}
& j(\underline{\beta})=1/2 {(\underline{\beta} - \underline{\beta}_b   )}^T B_\beta (\underline{\beta} - \underline{\beta}_b ) + \\
& 1/2 [\underline{y} - \hat{H}(\underline{x_r}, \underline{\beta})] ^T R^{-1} [\underline{y} - \hat{H}(\underline{x_r}, \underline{\beta})], 
\end{split}
\label{eq2}
\end{equation}
\noindent where $\underline{\beta}$ is the constant bias correction coefficients vector (generated for a week), $\underline{\beta}_b$ is the background bias correction coefficient vector, $B_\beta$ is the associated error covariance matrix for background bias correction coefficient vector, $\underline{y}$ is the observations, and $R$ is the observation error covariance matrix~\cite{Liu:2012:ImpactofAssimilating}. The objective of the 3DVar method is to minimize this cost function iteratively to increase the precision of the weather forecasts, and knowing the potential noises in measurements like radiance measurements could facilitate this process greatly by choosing the proper bias correcting coefficient for each predictor parameter used in this error function. In the next section, we present numerical results showing how 5G leakage from the n258 band could lead to perturbations in radiance measurements in the 23.8 GHz band, and hence impact the underlying temperature and rainfall forecasts that rely on such measurements.

\section{Numerical Results: Impact of 5G on Precipitation and Temperature Forecasts}
\label{sec4:NumericalResults}
A preliminary study was conducted to test the key hypothesis of this paper that the operation of 5G systems will affect the weather forecasting accuracy. For this case study, we chose a 12 hour forecasting period beginning at 12:00 UTC on Feb 05, 2008 covering the contiguous United States, due to readily available data with which to test the model. This simulation was designed to reproduce the incident of the Super Tuesday Tornado Outbreak~\cite{SuperTuesday}. On that day, 87 tornadoes were observed with a report of 57 casualties and \$1.2 billion loss, which was the second deadliest disaster in February in the US history \cite{chaney2010vulnerability}.

\begin{figure}[t!]
    \centering
    \subfigure{\centering \includegraphics[width=.47\linewidth]{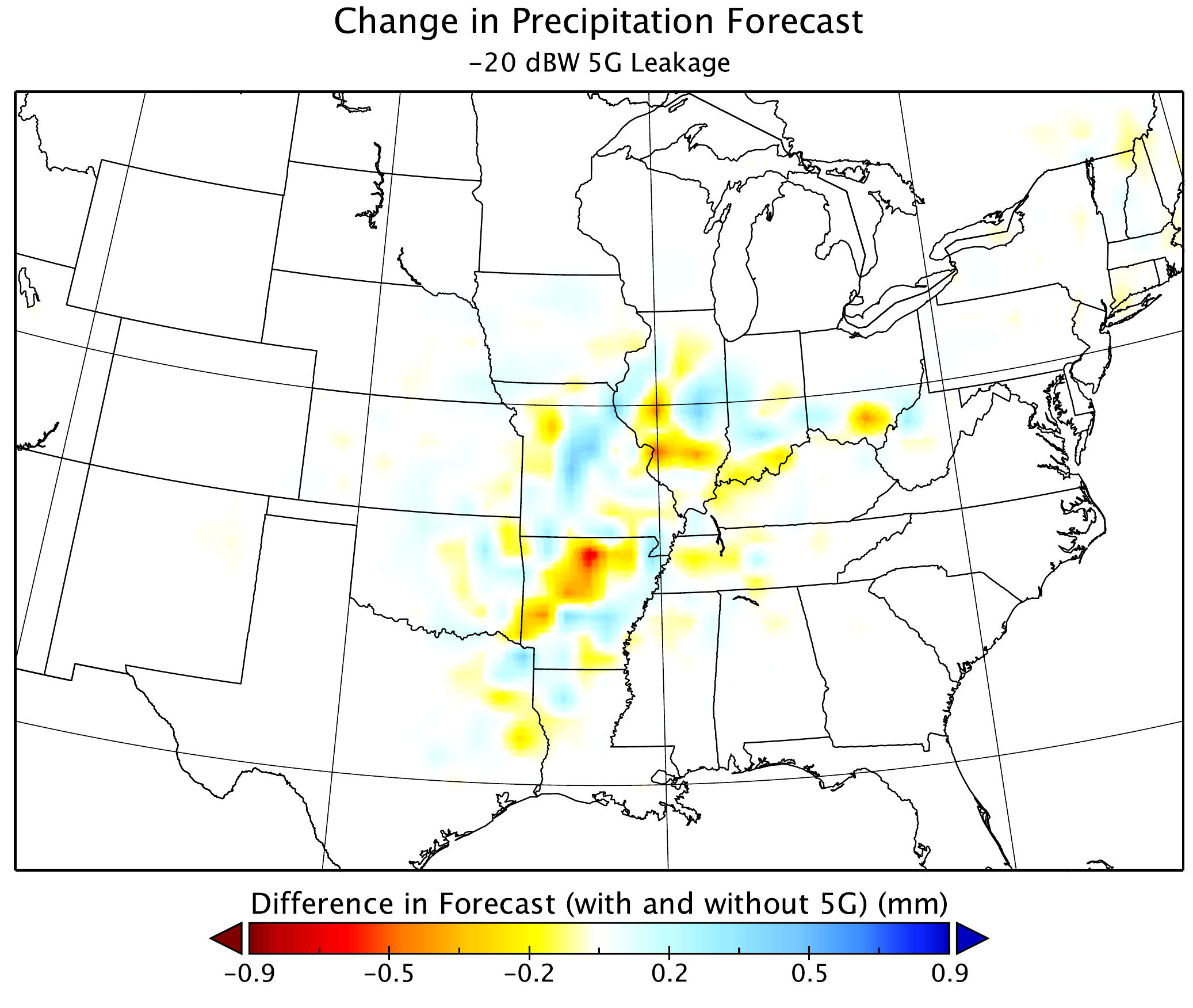} }
    \hfill
    \subfigure{\centering \includegraphics[width=.47\linewidth]{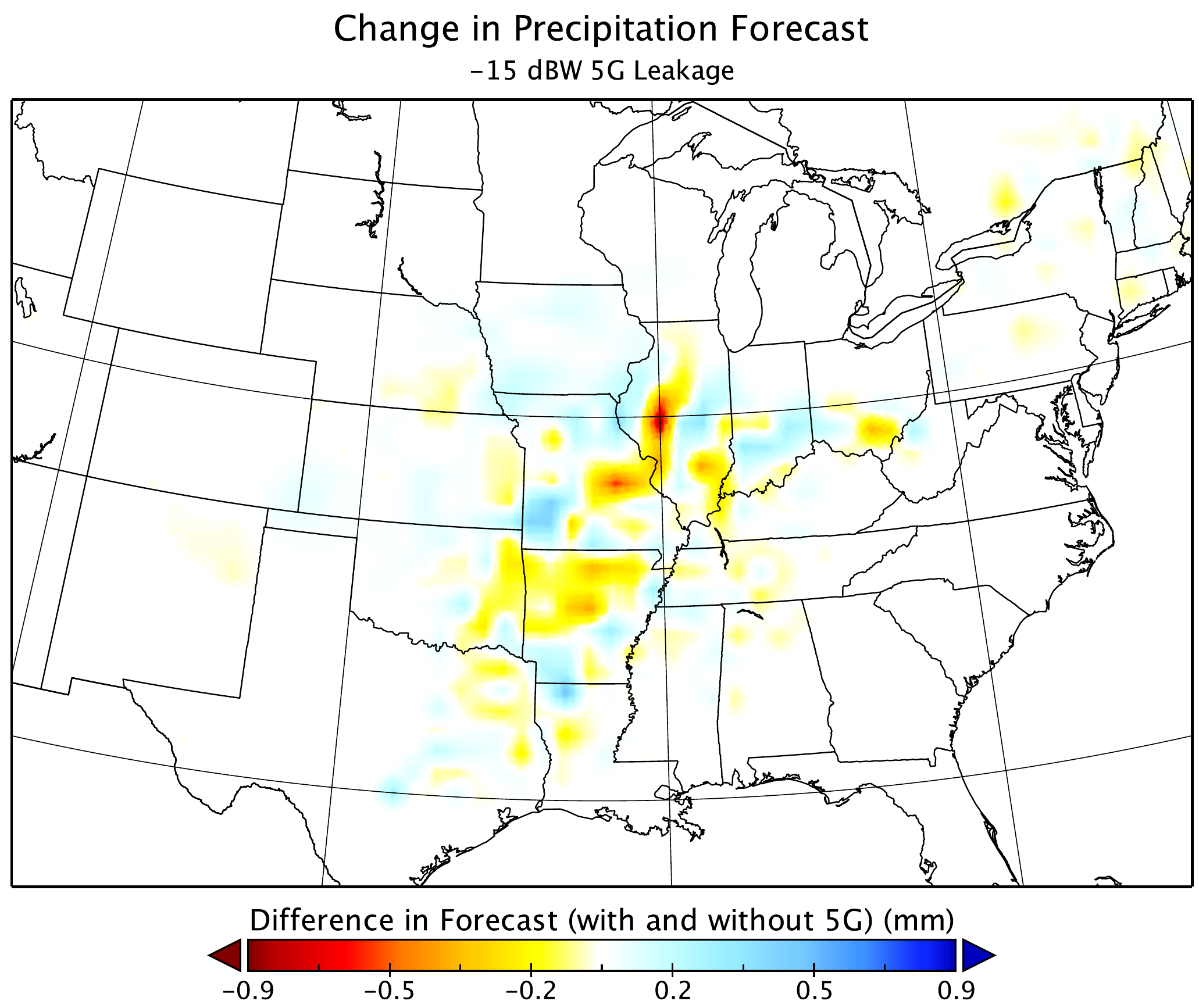}}
    \hfill
    \subfigure{\centering \includegraphics[width=.47\linewidth]{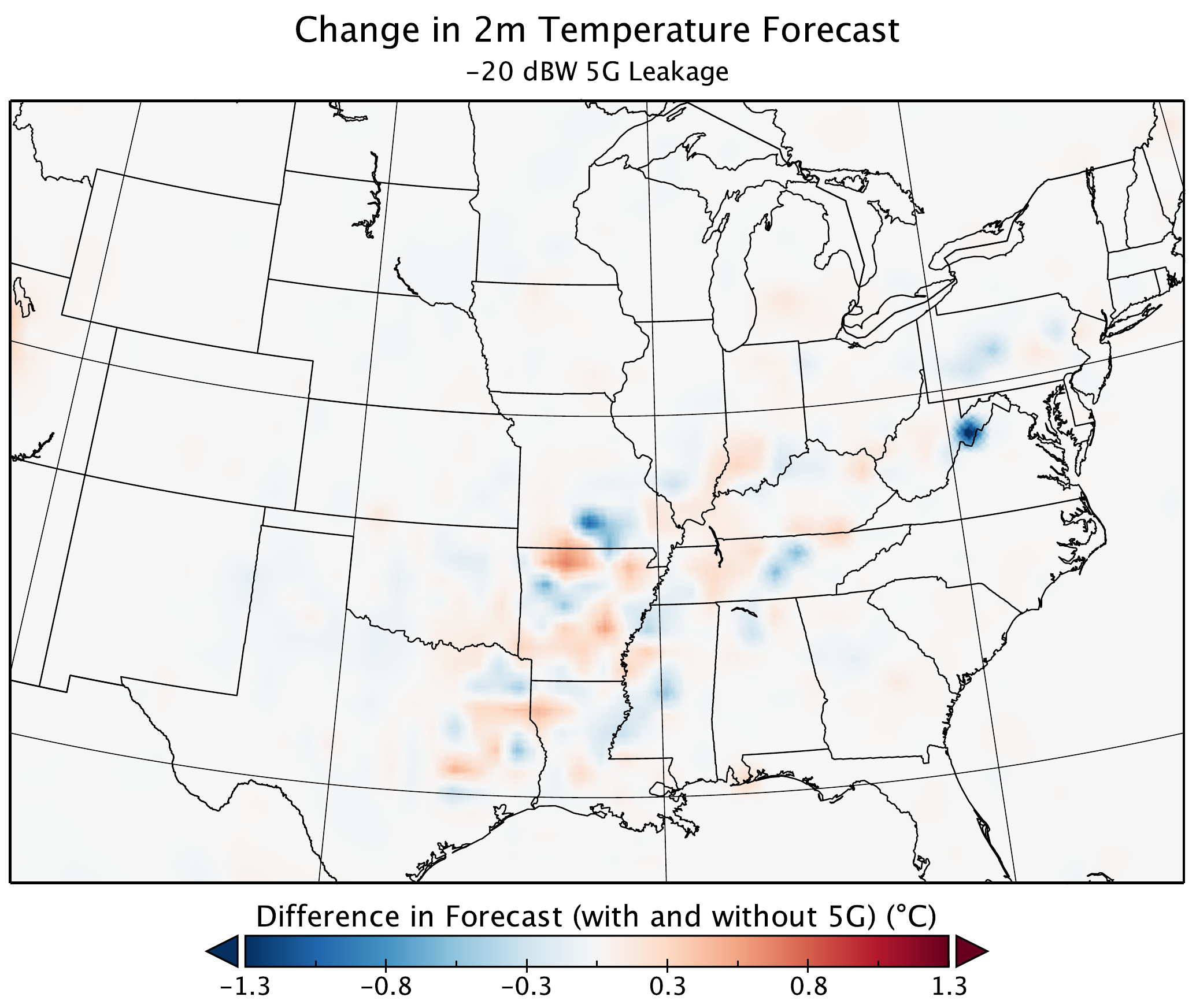} }
    \hfill
    \subfigure{\centering \includegraphics[width=.47\linewidth]{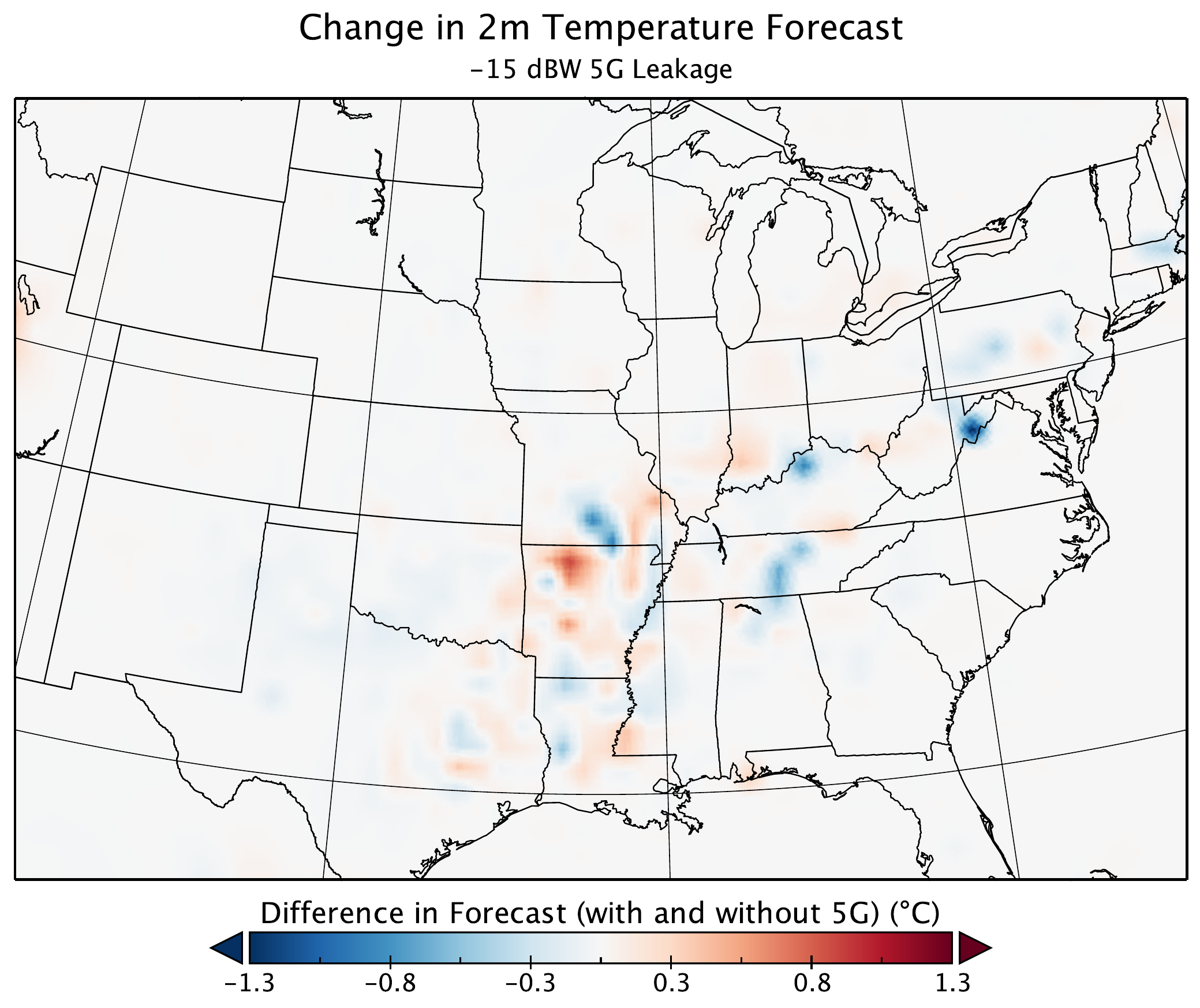} }
            \caption{The impact of 5G on precipitation and temperature forecasting.} \label{fig:T2}
            \vspace{-0.2in}
\end{figure}

We adopted the Weather Research and Forecasting Data Assimilation model (WRFDA), a data assimilation system built within the WRF software framework and used for application in both research and operational forecasting, to evaluate possible impacts that errors in the satellite radiance data from 5G may have on a widely used NWP model~\cite{Skamarock:2008:WRF, Barker}. The radiance data was collected from the Advanced Microwave Sounding Unit (AMSU) service. The first AMSU instrument was sent on the NOAA-15 satellite on 13 May 1998, which was joined by eight other satellites later. The service consists of 18 channels (15 with AMSU-A and 3 with AMSU-B) including the concerned 23.8 GHz band, which has a bandwidth of 270 MHz. The NOAA KLM series of satellites are flying in sunsynchronous polar orbits with constant altitudes ranging from 705 to 870 km and scanning a swath of 3.3\degree\ (48 km at nadir) to cover the earth's surface twice a day~\cite{AMSU,NOAAKLM2014}. The initial condition was provided by the testing data set of WRFDA \cite{testingdata}. A baseline case was created by assimilating the AMSU data at 12:00 UTC on Feb 5th, 2008 to update the initial and boundary conditions. WRF was employed to run with the assimilated starting point to proceed with the forecasting. 

To estimate the impact of 5G transmissions, we picked two levels of 5G leakage obtained from Fig.~\ref{NoiseFigure}: -20 dBW and -15 dBW. To assess the first-order magnitude of impact, we assumed that the leakage from 5G system was uniformly distributed over the earth's surface, i.e., the brightness temperature of the radiance data is increased by a constant level in the 23.8 GHz band (Channel 1). The two perturbed radiance data sets were then used to create two perturbed initial conditions. Boundary conditions were adjusted to match the perturbed assimilation. We ran WRF forecasting for 12 hours for each perturbed case respectively. Figure~\ref{fig:T2} shows the difference that 5G leakage causes to both the precipitation forecasts and the 2m-temperature (temperature at 2m above the earth's surface) forecasts compared to the baseline case. From the figure, it is noticeably clear that 5G can affect the meteorological forecasting up to 0.9 mm in precipitation and 1.3\,\celsius\ in 2m-temperature. This significant amount of impact motivates the need for a deeper investigation of the impact of 5G operation on weather forecasting and means to mitigate it. Therefore, it is imperative to develop refined models to improve our understanding of this impact and mechanisms to support the development of innovative schemes to allow co-existence of the two competing uses of the spectrum resource.

\section{Future Directions: Modeling and Mitigation}
\label{sec5:FutureDirections}
In this section, we discuss further directions for both modeling the impact of 5G on weather forecasting as well as mitigation. 

{\bf Improved Propagation Models for 5G Leakage and Induced Radiance:}
To better estimate the induced radiance due to the leakage of 5G signals at the passive radiometers on weather satellites, we need to use more detailed propagation models, in which the absorption loss of 5G signals as they pass through the atmosphere is considered. Knowing the absorption and transmittance coefficients of the atmosphere in any given geographical area, we can estimate the received power of 5G signals at the satellite sensors that are monitoring the atmosphere’s radiance. Note that both absorption and transmittance of any medium like the atmosphere are also a function of the passing signals' wavelengths and frequencies. In general, if we denote the absorption rate of the atmosphere for the signals with the wavelength of $\lambda$ with $\alpha_\lambda$, and its transmittance coefficient for such signals with $\tau_\lambda$, we always have $ \alpha_\lambda + \tau_\lambda =1$, which implies that the leaked 5G signal's energy will be either absorbed by the atmosphere or radiated out from it~\cite{Schwartz:2019:RadianceDataAssimilation}. 

{\bf Spatial Density, Elevation and Directionality:} The aggregate 5G leakage power into the 23.8 GHz band depends on the spatial density of the 5G transmitters in a given geographical area as well as the elevation from the earth's surface and their directionality. Hence, the density of outdoor base stations, UEs, drones and outdoor IoT devices will need to be assessed. In~\cite{Rappaport:2014:PathLossModels}, using field measurement based path loss models for 5G, it is shown that roughly three times more base stations (with cell radius up to 200 m) are required to accommodate 5G users as compared to existing 3G and 4G systems (with cell radius of 500 m to 1 km). 
Since the coverage radius associated with the pixel size of the AMSU sensor on weather satellites is 48 km~\cite{AMSU}, we could use 5G equipment density distribution models to evaluate the aggregate number of devices in a grid of 48 square km, and estimate their total energy leakage in a given angular direction that could reach the radiometers located at weather satellites. Given that there is variability in traffic demands across population centers, we should consider at least 2 classes of models corresponding to {\em metropolitan} and {\em rural} areas so that we can generate appropriate spatial density models by taking into account the number of 5G base stations, mobile devices of end-users and IoT devices in a given area. We should also characterize as function of time of day, these spatial densities to obtain dynamic spatial density models and then augment them with both typical elevation data on devices as well as models for directional transmissions. 

{\bf Device and Transmission Parameters:} Besides models for spatial density, elevation and directionality, we should additionally consider transmission parameters such as transmit power levels, specific sub-band occupancy, transmit modulation schemes and nonlinearity of power amplifiers in the transmitters, to evaluate the ACI caused to the 23.8 GHz band. Note that most of these parameters will vary depending on the type of device (base station, mobile device, or IoT). In~\cite{Singya:2017:MitigatingNLD}, it is shown that the nonlinear distortion of power amplifiers in 5G networks increases due to the use of large bandwidth in mmWave frequencies and operation near amplifier saturation. Hence, we anticipate that the impact of the leakage and the ACI to neighboring bands will be larger in mmWave bands.

{\bf  Cross-Layer Approaches for Mitigating 5G Impact on 23.8 GHz:}
To further mitigate 5G impact, we should design cross-layer PHY/antenna approaches to both spatially and spectrally mitigate 5G leakage. For spatial mitigation, using techniques such as direct antenna modulation (DM)~\cite{yao2004direct} seem promising. To spectrally mitigate 5G leakage, using enabling technologies like filtennas~\cite{luo2007filtenna} look promising. For example, to spatially direct 5G transmissions away from the direction of the passive weather sensors on satellites and also spectrally minimize the ACI into the 23.8 GHz band, we can use Direct Modulation Filtenna Array (DMFA), a mmWave filtenna (filtering antenna) array~\cite{luo2007filtenna} integrated with direct antenna modulation~\cite{yao2004direct} which enables sidelobe reduction for the antenna farfield radiation patterns.  
Moreover, we could integrate these PHY/antenna approaches into directional MAC and routing schemes with power control to meet the dual goals of mitigating 5G leakage and meeting desired QoS for 5G users, at the same time.

{\bf  Improved Weather Forecasting Algorithms:}
To enhance coexistence of 5G systems and reliable weather forecasting, we need to improve/update the data assimilation and NWP models used for weather forecasting, to adapt them based on the dynamics of 5G systems in time, space, and frequency. 
For example, Weather Research and Forecasting (WRF) DA (WRFDA) models could be utilized in several forecasting periods to precisely capture 5G leakage impacts on different weather events~\cite{Siegert2016,Suriano2017} by comparing the obtained snapshots for each period.
The preliminary results in section~\ref{sec4:NumericalResults} used 3DVar, which is a three-dimensional scheme. Since 5G leakage is dynamic in time, it is imperative to repeat the tests presented with 4DVar -- a DA scheme that adds the history of data to 3DVar to improve error correction~\cite{gauthier2007extension}. Moreover, to identify geospatial sensitivity of NWP to 5G leakage, stochastic modeling approaches could be implemented to systematically assess the sensitivity of forecasting models to synthesized radiance data disruptions resulting from 5G leakage. Traditional bias-correction methods primarily focus on removing cloud affects \cite{otkin2018nonlinear}, however, such schemes should be upgraded to be able to identify and remove 5G Leakage bias as well, thereby increasing the precision of weather forecasts. Overall, pre-processing of the radiance data before assimilation, passing the data through proper filters, and using 5G leakage-informed vector augmentation schemes, along with using machine learning approaches could all be helpful and effective in removing the radiance noise associated to 5G leakage.

\section{Conclusion}
\label{sec6:Conclusion}
Motivated by the concerns among the meteorological data forecasting community regarding 5G, in this paper, we investigated the impact of 5G mmWave leakage from the n258 band on the precision of weather forecasts that rely on observations of passive sensors (e.g. AMSU-A) operating at 23.8 GHz on weather satellites used to detect the amount of water vapor in the atmosphere. We studied the impact of 5G leakage on the accuracy of data assimilation based weather prediction algorithms by using
a first order propagation model to characterize the effect of the leakage signal on the brightness temperature (atmospheric radiance) and the induced noise temperature at the receiving antenna of the passive sensor (radiometer) on the weather observation satellite. We then characterized the resulting inaccuracies when using the Weather Research and Forecasting Data Assimilation model (WRFDA) to predict temperature and rainfall. For example, the impact of 5G leakage of -20dBW to -15dBW on the well-known Super Tuesday Tornado Outbreak data set, affects the meteorological forecasting up to 0.9 mm in precipitation and 1.3 °C in 2m-temperature. We also outlined future directions for both improved modeling of 5G leakage effects as well as mitigation using cross-layer antenna techniques coupled with resource allocation.

\bibliographystyle{IEEEtran}
\bibliography{main}
\end{document}